\documentclass[aps,prl,showpacs,preprintnumbers,amsmath,amssymb,nofootinbib,
epsf,superscriptaddress,
tightenlines]{revtex4}

\usepackage{graphicx}

\usepackage{graphicx}
\usepackage{amsmath}
\usepackage{bm}

\def\beqn{\begin{eqnarray}}
\def\eeqn{\end{eqnarray}}
\def\barr{\begin{array}}
\def\earr{\end{array}}
\def\btab{\begin{tabular}}
\def\etab{\end{tabular}}
\def\bite{\begin{itemize}}
\def\eite{\end{itemize}}
\def\bcen{\begin{center}}
\def\ecen{\end{center}}

\def\eq{\begin{equation}}
\def\ee{\end{equation}}
\def\nn{\nonumber}

\def\q2dagger{q_2\hspace{-0.35cm}/\;}

\begin{document}
\title{Forward Compton Scattering with weak neutral current: constraints from sum rules}
\author{Mikhail Gorchtein}
\affiliation{PRISMA Cluster of Excellence, Institut f\"ur Kernphysik, Johannes Gutenberg-Universit\"at, Mainz, Germany}
\email{gorshtey@kph.uni-mainz.de}
\author{Xilin Zhang}
\affiliation{Department of Physcs, University of Washington, Seattle, Washington, USA}
\email{xilinz@uw.edu}
\affiliation{Institute of Nuclear and Particle Physics and Department of Physics and Astronomy, Ohio University, Athens, Ohio, USA}
\affiliation{Fermi National Accelerator Laboratory, Batavia, Illinois, USA}
\begin{abstract}
We generalize forward real Compton amplitude to the case of the
interference of the electromagnetic and weak neutral current, 
formulate a low-energy theorem, relate the new amplitudes to the
interference structure functions and obtain a new set of sum rules.
We address a possible new sum rule that relates the product
of the axial charge and magnetic moment of the nucleon to the 0th
moment of the structure function $g_5(\nu,0)$. 
We apply the GDH and the finite energy sum rule for constraining the
dispersive $\gamma Z$-box correction to the proton's weak charge.
\end{abstract}
\maketitle


The study of Compton scattering within dispersion relation formalism has led to a derivation of the celebrated sum rules that model-independently relate low-energy properties of the nucleon to its excitation spectrum. Forward Compton amplitude that contains parity-conserving (PC) and parity-violating (PV) interactions is expressed in terms of scalar PC amplitudes $f,\,g$ and PV amplitudes $\tilde f,\,\tilde g$, see {\it e.g.} Refs. \cite{Bedaque:1999dh,Chen:2000mb},  
\beqn
T(\nu)&=&f(\nu)(\vec\varepsilon\,'^*\cdot\vec\varepsilon)
+g(\nu)i\vec\sigma\cdot[\vec\varepsilon\,'^*\!\times\vec\varepsilon]
+\tilde f(\nu)i\hat q\cdot[\vec\varepsilon\,'^*\!\times\vec\varepsilon]
+\tilde g(\nu)(\vec\sigma \hat q)(\vec\varepsilon\,'^*\cdot\vec\varepsilon),\label{eq:Comptonamplitude}
\eeqn
with $M$ the nucleon mass, $\vec\sigma$ the nucleon spin, $\hat q$ the unit vector pointing in the direction of the photon three momentum, and $\vec\varepsilon\,',\vec\varepsilon$ the final (intial) photon polarization vectors. We define the electromagnetic forward Compton amplitude as
\beqn
T_{\gamma\gamma}=\frac{i}{8\pi Me^2}\int d^4xe^{iqx}\langle
p|T\,j_{EM}^\mu(x)j_{EM}^\nu(0)|p\rangle\varepsilon_\mu\varepsilon\,'^*_\nu,
\eeqn
with $M$ the nucleon mass and $e$ related to the fine structure constant $\alpha_{em}=e^2/(4\pi)\approx1/137$. Only PC amplitudes $f,\,g$ are present in the electromagnetic case. The $\gamma Z$-interference forward Compton amplitude is normalized as
\beqn
T^{\gamma Z}=\frac{i\,\!\sin2\theta_W}{4\pi Me^2}\!\!\int\!\! d^4xe^{iqx}\langle
p|T\,j_{NC}^\mu(x)j_{EM}^\nu(0)|p\rangle\varepsilon_\mu\varepsilon\,'^*_\nu,
\eeqn
with $\theta_W$ the weak mixing angle. We focus on transverse $Z^0$ here, whereas the longitudinal component may be related to pion photo production through PCAC. The vector coupling of the $Z^0$ contributes to the amplitudes $f,g$ that already appeared in the electromagnetic case. To disambiguate we will use the superscript $\gamma Z$ for the interference case. The PV amplitudes $\tilde f, \tilde g$ arise from an interference of the electromagnetic current with the axial vector current, and correspond to nucleon spin-independent and nucleon spin-dependent contributions, respectively. Under crossing $\nu\to-\nu$ the amplitudes $f,\tilde g$ are even, while the amplitudes $\tilde f, g$ are odd. 

We wish to emphasize that although we consider a $Z^0$ boson in the final or initial state, the kinematics of the Compton process that we study here is such that the on-shell $Z^0$ cannot be produced since $q^2=0$. PV sum rules have been considered either in the Compton process $\gamma+N\to\gamma+N$ with hadronic PV effects \cite{Lukaszuk:2002ce,Kurek:2004ud} or Compton-like process $\gamma+\nu\to W^++e^-$, $\gamma+e^-\to Z^0+e^-$ and such, with an on-shell weak boson produced in the final state \cite{Altarelli:1972nc,Brodsky:1995fj}. In the process that we consider, the $Z^0$ may originate, {\it e.g.}, from neutrino or charged lepton scattering off the nucleon accompanied with a radiation of a real photon in the final state, as, e.g., virtual Compton scattering is accessed in a process $e^-+N\to e^-+N+\gamma$. In the context of, e.g.,  MiniBooNE \cite{MiniBNexcess} and other neutrino oscillation experiments, $\nu+N\to \nu+N+\gamma$ is an important background that has been addressed in phenomenological calculations \cite{zgammamesonints, Zhang:2012xn, Wang:2014nat, Harvey:2007rd, Hill:2009ek}. Forward $\gamma Z$ interference Compton amplitude enters the calculation of some electroweak corrections, in particular the dispersion $\gamma Z$-box correction to the weak charge of the proton \cite{Gorchtein:2008px,Sibirtsev:2010zg,Rislow:2010vi,Gorchtein:2011mz,Hall:2013hta} for the kinematics of the Q-Weak experiment currently under analysis \cite{Androic:2013rhu}. At present, the theory uncertainty is dominated by that due to the $\gamma Z$-box \cite{Gorchtein:2011mz}. PV and PC $\gamma Z$-interference structure functions are not constrained by experimental data, especially at low $Q^2$, and to perform calculations they have to be modeled applying symmetry and isospin structure assumptions to the electromagnetic data. The estimates for the theory uncertainty due to the $\gamma Z$-correction for the Q-Weak vary between 0.5\% \cite{Hall:2013hta} and 2.8\% \cite{Gorchtein:2011mz} of the Standard Model value of the proton's weak charge, and we investigate here, to what extent the sum rules for the interference Compton process constrain this calculation. 

The low energy limit of the Compton amplitudes is obtained by considering the ground state contribution, and it depends only on the nucleon mass $M$, charge (electric $e_N$, weak $Q_W^N$ and axial $g_A^N$) and magnetic moment (electromagnetic $\kappa_N$ or weak $\kappa_N^Z$). Parametrizing the next-to-leading order in the photon energy $\nu$ in terms of the polarizabilities, one obtains the low-energy expansion (LEX) up to order $\nu^2$: the well-known result  for the electromagnetic case \cite{GellMann:1954kc,Low},
\beqn
f(\nu)&=&-\frac{e_N^2}{4\pi M} +\frac{1}{e^2}(\alpha+\beta)^{\gamma\gamma}\nu^2+\dots,\nn\\
g(\nu)&=&-\nu\frac{(\kappa^\gamma_N)^2}{8\pi M^2} +\dots,
\eeqn
and the new results for the $\gamma Z$-interference,
\beqn
f^{\gamma Z}(\nu)&=&-\frac{e_NQ_W^N}{4\pi M} +\frac{1}{e^2}(\alpha+\beta)^{\gamma Z}\nu^2+\dots,\nn\\
g^{\gamma Z}(\nu)&=&-\nu\frac{\kappa^\gamma_N \kappa_N^Z}{8\pi M^2}+\dots,\nn\\
\tilde f(\nu)&=&0 +\frac{1}{e^2}\delta_1^{\gamma Z}\nu+\dots,\nn\\
\tilde g(\nu)&=&-\frac{g_A^N\mu_N}{4\pi M} +\frac{1}{e^2}\delta_2^{\gamma Z}\nu^2+\dots\label{eq:LEX}
\eeqn
Above, according to the definition of the interference Compton amplitude, we use 
$Q_W^p=1-4\sin^2\theta_W$, $Q_W^n=-1$, $g_A^p=-g_A^n=1.2701(25)$, 
$\kappa_p^Z=(1-4\sin^2\theta_W)\kappa^\gamma_p-\kappa^\gamma_n-\mu_s$, and 
$\kappa_n^Z=(1-4\sin^2\theta_W)\kappa^\gamma_n-\kappa^\gamma_p-\mu_s$. 
The strangeness contribution to the magnetic moment $\mu^s$, according to a recent global analysis of Ref. \cite{Armstrong:2012bi}, is $\mu_s=0.29\pm0.21$. The nucleon axial charge was taken without radiative corrections, and the strange quark contribution was neglected, since its effect is expected to be much smaller than the radiative corrections \cite{Armstrong:2012bi}.
Above, the nucleon magnetic moment was defined as $\mu_N=e_N+\kappa^\gamma_N$, and two new polarizabilities $\delta_{1,2}^{\gamma Z}$ were introduced. The optical theorem relates the imaginary parts of the forward amplitudes to the inelastic structure functions $F_{1,3}(\nu,Q^2),\,g_{1,5}(\nu,Q^2)$ taken in the limit $Q^2=0$:
\beqn
{\rm Im}\, f=\frac{1}{4M} F_1,\;\;\;\;\;\;\;
{\rm Im} \,g=\frac{1}{4M} g_1,\;\;\;\;\;\;\;
{\rm Im} \,\tilde f=\frac{1}{8M}F_3^{\gamma  Z},\;\;\;\;\;\;\;
{\rm Im}\, \tilde g=-\frac{1}{2M}g_5^{\gamma  Z}.
\eeqn
The amplitudes $f,\,g,\,\tilde f,\,\tilde g$ are analytic functions of complex energy and obey dispersion relations
\beqn
{\rm Re} \,f(\nu)&=& f(0)+\frac{\nu^2}{4\pi M}\int_{\nu_\pi}^\infty
\frac{d\nu'^2}{\nu'^2(\nu'^2-\nu^2)}F_1(\nu',0)\nn\\
{\rm Re} \,g(\nu)&=&\frac{\nu}{2\pi M}\int_{\nu_\pi}^\infty
\frac{d\nu'}{\nu'^2-\nu^2}g_1(\nu',0),\\
{\rm Re} \,\tilde f(\nu)&=& \frac{\nu}{4\pi M}\int_{\nu_\pi}^\infty
\frac{d\nu'}{\nu'^2-\nu^2}F_3^{\gamma Z}(\nu',0)\nn\\
{\rm Re} \,\tilde g(\nu)&=&\tilde g(0)-\frac{\nu^2}{2\pi M}\int_{\nu_\pi}^\infty
\frac{d\nu'^2}{\nu'^2(\nu'^2-\nu^2)}g_5^{\gamma Z}(\nu',0),\nn
\eeqn
where $\nu_\pi=m_\pi+m_\pi^2/2M$ is the first inelastic threshold due to pion production. The high-energy behavior of $F_1,\,g_5$ requires subtractions for $f,\tilde g$. 

These dispersion relations can now be evaluated for low energies $\nu\ll\nu_\pi$. Taylor-expanding the dispersion integrals in powers of $\nu^2$ and equating the coefficients in this expansion to the LEX of Eq. (\ref{eq:LEX}) the sum rules follow,
\beqn
(\alpha+\beta)^{\gamma\gamma\,,\gamma Z}&=&\frac{2\alpha_{em}}{M}\int_{\nu_\pi}^\infty
\frac{d\nu}{\nu^3}F_1^{\gamma\gamma\,,\gamma Z}(\nu,0) \label{baldin}\\
\kappa^\gamma_N\kappa^{\gamma,\,Z}_N&=&-4M\int_{\nu_\pi}^\infty
\frac{d\nu}{\nu^2}g_1^{\gamma\gamma,\,\gamma Z}(\nu,0), \label{gdh}\\
\delta_1^{\gamma Z}&=&\frac{\alpha_{em}}{M}\int_{\nu_\pi}^\infty
\frac{d\nu}{\nu^2}F_3^{\gamma Z}(\nu,0) \label{delta1sr},\\
\delta_2^{\gamma Z}&=&-\frac{4\alpha_{em}}{M}\int_{\nu_\pi}^\infty
\frac{d\nu}{\nu^3}g_5^{\gamma Z}(\nu,0) \label{delta2sr}.
\eeqn
Eqs. (\ref{baldin},\ref{gdh}) are Baldin \cite{Baldin} and Gerasimov-Drell-Hearn \cite{GDHsr} sum rules, respectively, and their straightforward generalization to the case of the $\gamma Z$ interference. Both sum rules were checked experimentally for the electromagnetic case \cite{GDH,OlmosdeLeon:2001zn} and the agreement was found to be better than 4\% for Baldin sum rule, and to be within 10\% for the GDH sum rule. The GDH sum rule was checked perturbatively in electroweak theory \cite{Altarelli:1972nc,Brodsky:1995fj,Dicus:2000cd}. Note that a GDH-like sum rule for PV Compton scattering was considered, {\it e.g.}, in Refs. \cite{Lukaszuk:2002ce,Kurek:2004ud} but in the context of hadronic parity violation, and not due to $\gamma-Z^0$ interference. 
The other two sum rules equate the PV polarizabilities $\delta_1$ and $\delta_2$ to the $1^{\rm st}$ moment of the structure function $F_3$ and $2^{\rm nd}$ moment of $g_5$, respectively, and both integrals are certainly convergent. 

Finally, the finite energy sum rule (FESR) for the amplitudes $f$ and $\tilde g$ results from extracting Regge-behaved part $f^R(\tilde g^R)$ of the amplitude $f(\tilde g)$ explicitly, and writing a dispersion relation for the difference $f-f^R$ and $\tilde g-\tilde g^R$. At the asymptotically high energy such an amplitude can be at most a constant that is denoted by $C_\infty(\tilde C_\infty)$, and one obtains a dispersion representation for this constant (the $J=0$ pole) \cite{Damashek:1969xj},
\beqn
C_\infty=-\frac{e_N^2}{4\pi M}-\frac{1}{2\pi M}\int_{\nu_{thr}}^N\frac{d\nu}{\nu}[F_1(\nu,0)-F_1^R(\nu,0)].
\label{eq:FESRgaga}
\eeqn
Above, $F_1^R=\sum_{i}c_i\nu^{\alpha_i}$ with $\alpha_i$ strictly positive. The leading high-energy behavior is described by the Pomeron with $\alpha_P\approx1.09$ and the $f_2$-trajectory exchange with $\alpha_{f_2}\approx0.5$, and was obtained from a Regge fit at $\nu\geq N\approx2$ GeV \cite{gorchtein_FESR}. Note that due to different normalization of the Compton amplitude, $C_\infty$ in Eq. (\ref{eq:FESRgaga}) differs from that in \cite{Damashek:1969xj,gorchtein_FESR} by a factor $(4\pi\alpha_{em})^{-1}$. Quite analogously we obtain for the interference PC amplitude,
\beqn
C_\infty^{\gamma Z}=-\frac{e_NQ_W^N}{4\pi M}
-\frac{1}{2\pi M}\int_{\nu_\pi}^N\frac{d\nu}{\nu}[F_1^{\gamma Z}(\nu,0)-F_1^{\gamma Z,\,R}(\nu,0)],
\label{FESRgaZ}
\eeqn
and for the interference PV amplitude,
\beqn
\tilde C_\infty^{\gamma Z}=-\frac{g_A^N\mu_N}{4\pi M}+\frac{1}{\pi M}\int_{\nu_{thr}}^N\frac{d\nu}{\nu}[g_5^{\gamma Z}(\nu,0)-g_5^{\gamma Z,\,R}(\nu,0)].
\label{eq:FESRg5}
\eeqn

It is necessary to stress that the FESR of Eq. (\ref{eq:FESRg5}) is based on the assumption that $g_5^{\gamma Z,\,R}$ diverges at the infinity as $\nu^\alpha$ with $\alpha>0$. Should this assumption not hold, and the structure function $g_5$ decrease at high energies, then an unsubtracted sum rule would have to be postulated,
\beqn
{g_A^N\mu_N}=4\int_{\nu_{thr}}^\infty\frac{d\nu}{\nu}g_5^{\gamma Z}(\nu,0).
\label{eq:SRg5}
\eeqn

To assess these options, we examine the high-energy asymptotics of $g_5$ more closely. At high energy and in the Regge framework , the structure accompanying $g_5$ may come about due to an exchange of an axial vector meson. Possible lowest mass candidates are $h_1(1170)$, $b_1(1235)$ , and $a_1(1260)$.  Due to lack of sufficient higher spin states for these channels the Chew-Frautschi plot for these trajectories is not fully constrained, and we will give a range for the intercept of these trajectories. The upper limit stems from relating an axial vector to the pion trajectory, thus $\alpha_0\approx-\alpha'm_\pi^2\approx-0.02$. The lower limit results from a linear extrapolation $\alpha_0^M=1-\alpha'm_M^2$ that range from $-0.1$ to $-0.4$ for the three candidates. We refer to two recent studies of the properties of Regge trajectories in \cite{Masjuan:2012gc} and \cite{Ford}. The latter Ref. includes an analysis of polarized $NN$ data up to high energy. Our simple estimate is in line with these two studies. Other works, {\it e.g.} \cite{oai:arXiv.org:1007.3140}, use $\alpha_{b_1}(0)\approx0.5$ which would require a much smaller Regge slope or a substantial nonlinearity of the respective trajectory. 
Apart from nucleon and meson scattering, the high-energy behavior of $g_5$ enters parametrizations of the polarized quark PDFs. Ref. \cite{oai:arXiv.org:0904.3821} obtains for the two lightest flavors, $g_5^u\sim(\Delta u-\Delta\bar u)(x\to0)\sim x^{-0.308}$ and $g_5^d\sim(\Delta u-\Delta\bar u)(x\to0)\sim x^{-0.836}$. In terms of possible Regge exchanges, this may suggest, upon assuming a universal Regge slope, an existence of two degenerate axial vector meson trajectories (isoscalar and isovector) realized as particles with masses below $\rho(770)$. No such states have been observed or predicted. On the other hand, the most recent analysis of polarized DIS data performed in Ref. \cite{Jimenez-Delgado:2013boa} obtains the low-$x$ behavior of the polarized valence PDF's for which an unsubtracted dispersion integral converges. The situation remains inconclusive, as existing polarized DIS data do not unambiguously constrain the high-energy asymptotics of $g_5(\nu,0)$. 

As an informative check, we consider the contribution of the $\Delta(1232)$ resonance to the isoscalar (in the isovector combination it drops out) sum rule of Eq.(\ref{eq:SRg5}). Accounting for the dominant magnetic $\gamma N\Delta$ coupling $c_{1\Delta}$ and the axial $Z^0N\Delta$ coupling $h_A$ (we refer the reader to Refs. \cite{Serot:2011yx,Serot:2012rd} and references therein for details), we obtain,
\beqn
g_A(\mu_p-\mu_n)=-\frac{32}{9}h_Ac_{1\Delta}\left(M+M_\Delta+\frac{M_\Delta^2-M^2}{2M_\Delta}\right)
\frac{1}{\pi}
\int\limits_{\nu_\pi}^\infty\! d\nu\,{\rm Im}\left[\frac{1}{W^2-M_\Delta^2+iM_\Delta\Gamma_\Delta(W)}\right],
\eeqn
with $W^2=M^2+2M\nu$ the invariant mass of the intermediate hadronic state. Using the values for the $\Delta$ parameters from Refs. \cite{Serot:2011yx,Serot:2012rd}, $h_A=1.40,\;c_{1\Delta}=1.21$ and treating the $\Delta(1232)$ as a narrow resonance leads to r.h.s.$\approx16/9h_Ac_{1\Delta}(1+M_\Delta/M)\approx7.79$; using the experimental width and accounting for its energy dependence reduces the result to $5.93$, to be compared to the l.h.s.$g_A(\mu_p-\mu_n)=5.96$. The agreement is remarkable. The only viable Regge exchange, $h_1(1170)$ seems to be consistent with negative intercept in all analyses known to us, while $a_1,\,b_1$ do not contribute being isovectors. Further contributions to the sum rule still have to be incorporated, such as non-resonant $\pi N$ contributions, and higher resonance states. \\ 

{\bf Sum rule for $\delta_2^{\gamma Z}$}\\
Our numerical estimate in the model with the narrow $\Delta(1232)$ leads to 
$\delta_2^{\gamma Z,\,p}=\delta_2^{\gamma Z,\,n}\approx-2.0\times10^{-3}{\rm fm}^3.$
For comparison, the proton's electric polarizability is about half that size, $\alpha_E^p=(1.12\pm0.04)\times10^{-3}$ fm$^3$ \cite{PDG}. A more realistic estimate of the polarizability should include, e.g. the threshold pion production mechanism that is expected to be important numerically due to $1/\nu^3$ weighting under the integral.\\

{\bf Sum rule for $\delta_1^{\gamma Z}$}\\
The low-energy limit of the amplitude $\tilde f$ has been estimated in Ref. \cite{Harvey:2007rd} upon introducing an anomalous $\gamma Z^0\omega$ vertex, and in Ref. \cite{Hill:2009ek} with the $\Delta(1232)$ isobar. The latter mechanism turns out to be numerically more important. Our numerical estimate in the model with the narrow $\Delta(1232)$, 
$\delta_1^{\gamma Z,\,p}=\delta_1^{\gamma Z,\,n}\approx4.5\times10^{-3}{\rm fm}^2$, is consistent with that of Ref. \cite{Hill:2009ek}. It has been argued \cite{Harvey:2007rd} that this polarizability can induce an effective $\gamma\nu\bar\nu$ interaction that may provide an additional channel for energy loss from neutron stars.\\ 

\begin{figure}
\centering
  \begin{tabular}{@{}cc@{}}
    \includegraphics[width=.45\textwidth]{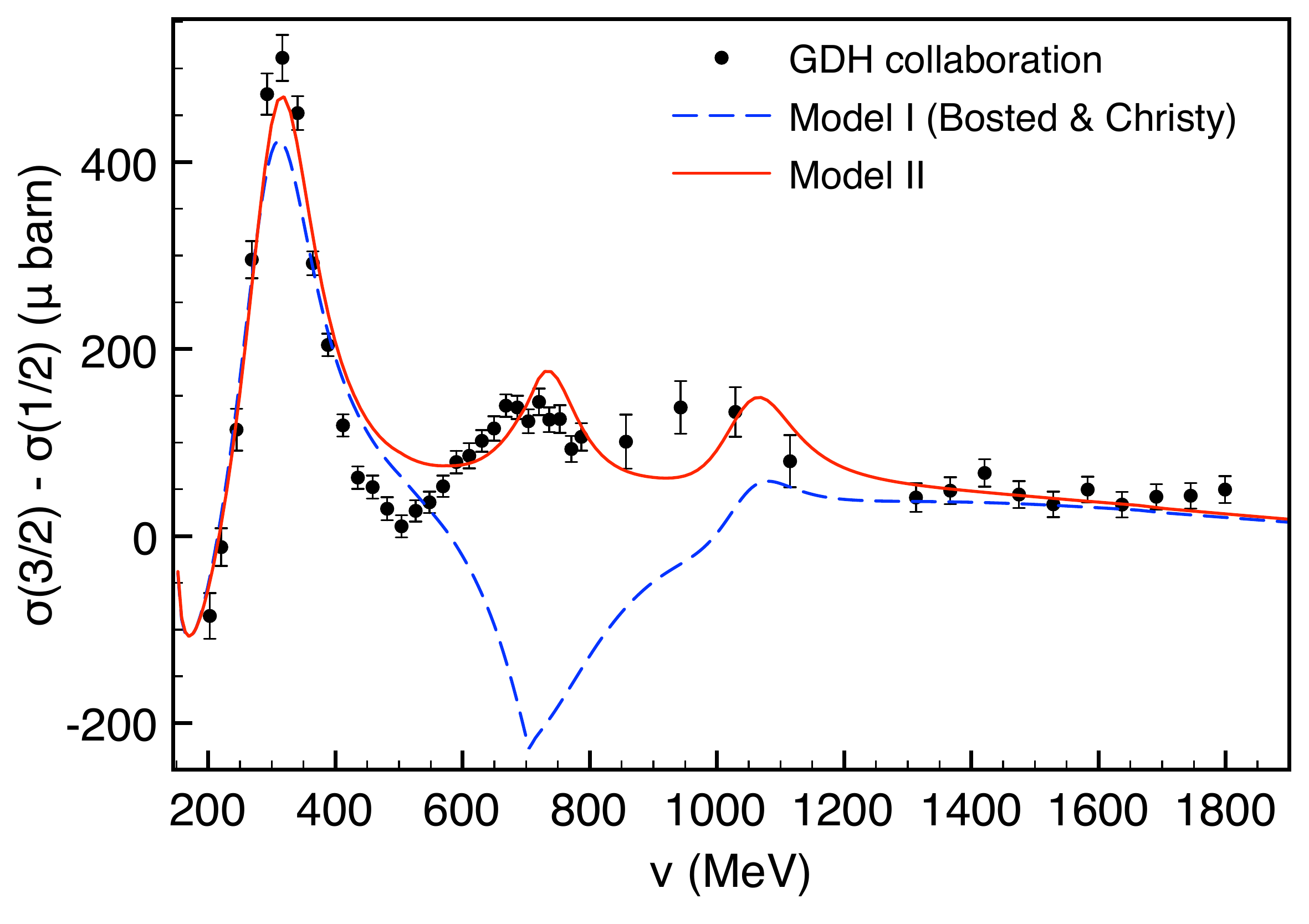} &
    \includegraphics[width=.45\textwidth]{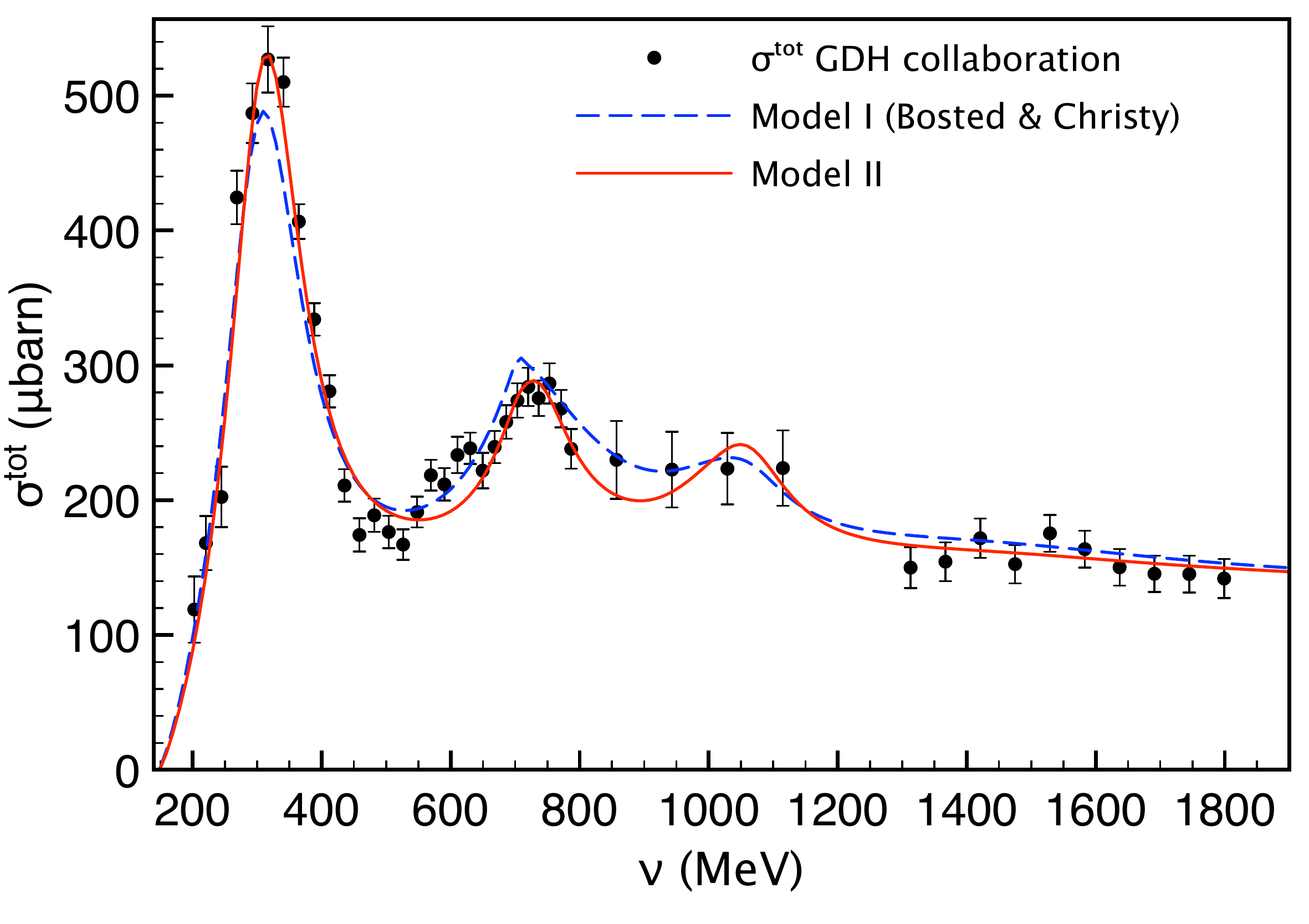} 
  \end{tabular}
\caption{Left panel: helicity difference $\sigma_{3/2}-\sigma_{1/2}$ photoabsorption cross section from the original parametrization of Ref. \cite{bosted_p} (dashed blue line) and the modified one adjusted in accord with the PDG \cite{PDG} (solid red line) in comparison with data by GDH collaboration \cite{GDH}. Right panel: same for the total photo absorption cross section.
}
\label{fig:gdh}
\end{figure}

{\bf GDH sum rule and the parametrization of resonance data}\\
A parametrization of the inelastic structure functions $F_{1,2}^{\gamma\gamma}$ on the proton target in the resonance region has been proposed by Christy and Bosted in Ref. \cite{bosted_p}. Consequently, this parametrization was used to predict the interference structure functions $F_{1,2}^{\gamma Z}$ that enter the calculation of the dispersive $\gamma Z$-box correction to the weak charge of the proton in the kinematics of the QWEAK experiment \cite{Androic:2013rhu}. The procedure involves a rotation of the transition helicity amplitudes for individual resonances in the weak isospin space, and is described in full detail in Ref. \cite{Gorchtein:2011mz}. It is based on the conservation of the vector current (CVC) and on the identification of quantum numbers of the resonances. The latter was taken from the original parametrization of Ref. \cite{bosted_p}. We will assess this identification with the use of the GDH sum rule. The parametrization of Ref. \cite{bosted_p} features two close resonances in the second resonance region, $S_{11}(1535)$ and $D_{13}(1520)$, of which the former one dominates carrying $\approx90$\% of the strength in the sum of the two for the total cross section.
The commonly accepted picture \cite{PDG} is nearly opposite, 
and this difference can be disentangled with the GDH sum rule: $S_{11}(1535)$ being a $J=1/2$ resonance cannot be excited in the $A_{3/2}$ channel, thus its contribution to the GDH sum rule is strictly negative (the spin structure function $g_1$ is related to the helicity-dependent photo absorption cross section as $g_1(\nu,0)=\frac{M\nu}{2\pi e^2}[\sigma_{1/2}-\sigma_{3/2}]$). 
Similarly, we consider $P_{11}(1440)$,  $S_{11}(1650)$ and  $F_{15}(1680)$. 
\begin{table}[h]
  \begin{tabular}{c|c|c|c|c|c}
\hline
& $S_{11}(1535)$ & $D_{13}(1520)$ & $F_{15}(1680)$ & $S_{11}(1650)$ 
& $P_{11}(1440)$ 
\\
\hline
$A_T^I(0)$ & 
6.335 & 0.603 & 2.330 & 1.979 & 0.0225 
\\
\hline
$A_T^{II}(0)$ & 
3.3 & 3.5 & 3.1 & 2.0 & 2.422 
\\
\hline
 \end{tabular}
\caption{Values of the parameter $A_T$ for the 5 of 7 resonances used in the fit of Ref. \cite{bosted_p}. 
$A_T^I(0)$ stand for the original values and is referred to as Model I in the text.
$A_T^{II}(0)$) shows the values modified in accord with the PDG as described in the text and referred to as Model II.}
\label{tab1}
\end{table}
We display in Table \ref{tab1} how resonance parameters should be changed to be in agreement with the helicity difference cross section $\sigma_{3/2}-\sigma_{1/2}$ without affecting the description of the data for the total cross section, see Fig. \ref{fig:gdh}. The curves are compared to the data from Ref. \cite{GDH} that with certainty exclude the blue dashed curve (Model I). The red solid curve (Model II) compares favorably to the data. 
Each curve can be used to evaluate the r.h.s. of the GDH sum rule. Model I leads to $\kappa_p^2\approx0.9$, whereas Model II leads to $\kappa_p^2\approx3.28$, a result close to the sum rule value, $\kappa_p^2=1.793^2\approx3.215$. Note that for this evaluation we supplemented the threshold region with the non-resonant background contribution from MAID \cite{MAID} that gives a sizable negative contribution. This contribution, being the helicity-difference cannot be directly obtained from the parametrization of Ref. \cite{bosted_p} which only deals with the total cross section. 

We evaluate the PV analogue of the GDH sum rule with the isospin-rotated cross sections, and compare it to $\kappa_p^\gamma\kappa_p^Z$. The evaluation with Model I parametrization leads to $\kappa_p^\gamma\kappa_p^Z\approx2.247$, and that with Model II gives $\kappa_p^\gamma\kappa_p^Z\approx3.615$, to be compared to the l.h.s. $\kappa_p^\gamma\kappa_p^Z=(1-4s^2_w)(\kappa_p^\gamma)^2-\kappa_n^\gamma\kappa_p^\gamma\approx3.666$. The Model II is consistent with both sum rules. In this evaluation we neglected the strangeness contribution, and we did consistently so both in the l.h.s. low-energy coefficient $\kappa^\gamma_p\kappa^Z_p$ and in the r.h.s. integral. The uncertainty of the strange magnetic moment $\mu_s=0.29\pm0.21$ \cite{Armstrong:2012bi} can be used to assess the uncertainty of the l.h.s. of the GDH sum rule as $\delta\mu_s\kappa_p=\pm0.377$, significantly smaller than the deviation of the evaluation with Model I from the sum rule value. 

Now, we are in the position to update the value and the uncertainty of the dispersion evaluation of the 
resonance contribution to the Re$\,\Box_{\gamma Z}^V$ correction to the QWEAK measurement \cite{Gorchtein:2011mz} . The sum of the resonance contributions to Re$\,\Box_{\gamma Z_V}$ with the resonance parametrization of Model I amounted to Re$\,\Box_{\gamma Z_V,\,Res.}^{Mod.\,I}=2.24^{+0.53}_{-0.43}\cdot10^{-3}$. After the modification described above we arrive at Re$\,\Box_{\gamma Z_V,\,Res.}^{Mod.\,II}=2.23^{+0.28}_{-0.23}\cdot10^{-3}$, with the uncertainty sizably reduced. \\

{\bf Potential of FESR for constraining the uncertainty of the
  dispersive $\gamma Z$-box calculation}\\
The modification of resonance parameters described in the previous section also affects the description of the total cross section. The right panel of Fig. \ref{fig:gdh} displays the comparison of the two models and the data on $\sigma_{tot}$ obtained by the GDH collaboration \cite{GDH}. To improve the description of the data, we adjusted the value of the width of the $\Delta(1232)$ to the standard value of 120 MeV \cite{PDG}, while the fit of \cite{bosted_p} features a slightly larger width, 136 MeV. We will now feed these two parametrizations in the finite energy sum rule of Eq.(\ref{eq:FESRgaga}), and in its  $\gamma Z$-interference analogue of Eq.(\ref{FESRgaZ}).
The numerical evaluation of Eq. (\ref{eq:FESRgaga}) with resonance contributions from Model II and the background parametrization from \cite{gorchtein_FESR,Gorchtein:2011mz} leads to 
$C_\infty=-12.2\,\mu{\rm b \,GeV}$,
which agrees reasonably well with the extraction of the $J=0$ pole in \cite{Gorchtein:2011mz}, 
$C_\infty=-8.2\pm3.8\,\mu{\rm b \,GeV}$. 
To summarize the two evaluations, 
\beqn
C_\infty=-10.2\pm3.8(stat.)\pm2.0(syst.)\,\mu{\rm b \,GeV}, 
\label{eq:j=0}
\eeqn
where we estimate the systematical uncertainty by averaging over the two evaluations. The parametrization of Model I gives a larger result, $C_\infty=-18.3\,\mu$b GeV. Next we evaluate the $\gamma Z$-interference analog of the $J=0$ pole, Eq. (\ref{FESRgaZ}) using the isospin rotation as described in  \cite{Gorchtein:2011mz}. 
Model II leads to 
\beqn
C_\infty^{\gamma Z}=28.5\pm22.0(back.)^{+10.1}_{-8.5}(res.)\,\mu{\rm b \,GeV},
\label{eq:j=0gZpdg}
\eeqn
where the first uncertainty is due to the isospin structure of the background, and the second one due to that of the resonances.
Note that while the latter is obtained from data (analyzed by PDG) and can be considered reliable, the former uncertainty is not as solid. 
It is seen that the Model II evaluation may in principle be used to constrain the background contribution since the uncertainty of the latter dominates over that due to resonances. To do that, the information about the l.h.s. of Eq. (\ref{eq:j=0gZpdg}) is necessary. It has been argued in the literature that the $J=0$ pole, if exists, should be due to an effective two-photon-quark coupling. Then, knowing the $J=0$ pole for Compton scattering, Eq. (\ref{eq:j=0}) we can try to model the $\gamma Z$-interference $J=0$ pole. If at asymptotically high energy the dominant picture is a symmetric collection of quarks (SU(6) symmetry) one would expect
$C_\infty^{\gamma Z}/C_\infty\sim2\sum_{q}g_V^q e_q/\sum_{q}e_q^2=9/5-4s_W^2\approx0.85$,
whereas for the case where the $J=0$ pole is due to coupling to valence quarks one would have
$C_\infty^{\gamma Z}/C_\infty\sim2 (2g_V^u e_u+g_V^d e_d)/(2e_u^2+e_d^2)=5/3-4s_W^2\approx0.71$.
Yet another possibility is, as early evaluations of the FESR and the $J=0$ pole obtained, that the $J=0$ pole is equal to the Thomson term. In this case, we would have
$C_\infty^{\gamma Z}/C_\infty=Q_W^p\approx0.075$.
These estimates indicate that the $\gamma Z$-interference $J=0$ pole is likely to be somewhat smaller than the electromagnetic one, and to have the same sign. We can then assume that, conservatively,
\beqn
C_\infty^{\gamma Z}=-5.1\pm5.1\,\mu{\rm b \,GeV}.
\label{eq:j=0gZest}
\eeqn
\indent
A comparison with the evaluation of Eq. (\ref{eq:j=0gZpdg}) suggests that for the two to agree, the background contribution should be taken at its lower range, suggesting that the model of Ref. \cite{Gorchtein:2011mz} is likely to overestimate the interference structure functions $F_{1,2}^{\gamma Z}$ at $Q^2=0$. The recent measurement of the PV asymmetry in the resonance region on the deuteron \cite{Wang:2013kkc} observed that models tend to overshoot the data in the $\Delta(1232)$ region by 25-30\%, although the disagreement is not striking because of large uncertainties. Our analysis implies that this discrepancy may need to be taken seriously, since another physical constraint from FESR suggests the same behavior.\\

In summary, we derived a set of sum rules for forward Compton scattering generalized to the case of electromagnetic-weak neutral current interference. Along with a straightforward generalization of the GDH, Baldin and finite energy sum rules, we proposed a new sum rule that relates the product of the nucleon's axial charge and magnetic moment to an integral over the parity-violating structure function $g_5$. We analyzed the Regge asymptotics of that amplitude and found that currently, no solid statement about the convergence of this sum rule can be made. A model calculation for the isoscalar sum rule (its convergence is more reliable from the Regge stand point) with the $\Delta(1232)$ resonance leads to a very good agreement. This sum rule deserves further study: if confirmed it may give a constraint on the low-$x$ behavior of the polarized PDF's parametrizations. We showed that accounting for GDH and finite energy sum rules for electromagnetic and electroweak Compton amplitudes can help constraining parametrizations of inclusive electromagnetic and interference structure functions. The latter are important for calculating nucleon structure-dependent electroweak corrections to precision low-energy tests, {\it e.g.}, the proton's weak charge measurement. 

\acknowledgments{
M.G. is grateful to M. Vanderhaeghen, V. Pascalutsa, P. Masjuan and H. Spiesberger for useful discussions and comments. 
The work of M.G. was supported by the Deutsche Forschungsgemeinshaft DFG through the Collaborative Research Center ``The Low-Energy Frontier of the Standard Model'' (SFB 1044) and the Cluster of Excellence ÒPrecision Physics, Fundamental Interactions and Structure of MatterÓ (PRISMA). X.Z. thanks D. Phillips, L. Alvarez-Ruso, G. Zeller, and G. Miller for interesting discussions, and acknowledges support from the US Department of Energy under grant DE-FG02-93ER-40756, and from Fermi National Accelerator Laboratory under intensity frontier fellowship.
}

\end{document}